\documentclass[submission,copyright,creativecommons]{eptcs}
\providecommand{\event}{AMMSE 2011} 

\usepackage{breakurl}             

\usepackage{amsmath}
\usepackage{amsfonts}
\usepackage{amssymb}

\usepackage{amscd}
\usepackage{url}
\usepackage{times}
\usepackage{AMMALanguages}
\usepackage{graphicx}

\usepackage[latin1]{inputenc}
\usepackage{latexsym}
\usepackage{alltt}
\usepackage{shortvrb}  
\usepackage{dsfont}    

\newcommand{\figeps}[3]{ 
    \begin{figure}[h!tb] \begin{center}
       \includegraphics[height=#2mm]{Images/#1}
        \caption{#3}
        \label{fig:#1}\end{center}
    \end{figure}
}

\newcommand{\msgloss}[1]{}

\newcommand{\code}[1]{{\small\textsf{#1}}}

\title{Formal Visual Modeling of Real-Time Systems in
  e-Motions: Two  Case Studies}

\author{Francisco Dur\'an
\institute{
           Universidad de M\'alaga, Spain}
\and Peter Csaba \"Olveczky
\institute{
           University of Oslo, Norway}
\and Jos\'e E. Rivera
\institute{
           Universidad de M\'alaga, Spain}
}

\def\titlerunning{Formal Visual Modeling of Real-Time Systems in
  e-Motions: Two  Case Studies}
\def\authorrunning{F. Dur\'an,  P. C. \"Olveczky,
  and J. E. Rivera}

\begin{document}
\maketitle
 
\begin{abstract}
e-Motions is an Eclipse-based visual timed model transformation framework with
a Real-Time Maude semantics that supports the usual  Maude formal analysis methods,
including simulation, reachability analysis, and LTL model checking. 
e-Motions is characterized by a novel and powerful set of constructs for expressing
timed behaviors. 
In this paper we illustrate the use of these constructs --- and
thereby implicitly investigate their suitability to define 
real-time systems in an intuitive way --- to define and formally analyze two prototypical 
and very different real-time
systems: (i) a simple round trip time protocol for computing the time it takes a message
to travel from one node to another, and back; and (ii) the EDF scheduling
algorithm. 
\end{abstract}

\section{Introduction}
\label{sec:intro}

\emph{Model-driven engineering} (MDE) is becoming a widely
accepted approach for developing complex distributed applications.
MDE advocates the use of models as the key artifacts in all phases
of system development, from system specification and analysis to
 implementation. For this purpose, it makes use mainly of two
technologies: \emph{Domain-specific modeling languages} (DSMLs) and
\emph{model transformations}. 

DSMLs are used to represent the various
facets of a system in terms of models. Such languages  support
higher-level abstractions than general-purpose modeling languages, and
are closer to the problem domain than to the implementation domain. 
The \emph{abstract syntax} of a DSML defines its domain concepts
and their structural relationships. In MDE, the abstract syntax of a
DSML 
is usually defined by  a metamodel, which can 
be seen as a UML class diagram  that  describes the concepts
 of the language and the structuring
 rules that constrain the model elements  and their combinations.
The \emph{concrete syntax} of such a DSML defines the actual notation
used to represent the domain concepts, and is typically given as a
mapping
from the metamodel elements to a textual or (for visual languages)
graphical notation. 

In MDE, the \emph{behaviors} of a DSML can be defined as \emph{in-place model
transformations}, which define the possible dynamic evolutions of
valid models in the language. There are several
in-place model transformations approaches for 
specifying the behavior of a DSML
(see~\cite{Rivera-Guerra-deLara-Vallecillo:2008} 
for a brief survey). 

Real-time embedded systems, such as, e.g., embedded medical devices
and automotive and  avionics systems, are both hard to design
correctly, since subtle timing issues impact system correctness, yet
are safety-critical systems, whose failures could result in significant
losses to human lives and/or valuable assets. Therefore, the applications of MDE 
to such real-time systems should be supported by automated formal
analyses. 
To the best of our knowledge, there are two model transformation
frameworks for real-time systems within the Eclipse framework that
support both reachability analysis and LTL model checking. They both
have a Real-Time Maude~\cite{tacas08} semantics and can be 
formally 
analyzed in Maude~\cite{maude-book}.

One such framework is the timed extension of the MOMENT2 model
transformation framework defined by Boronat and \"Olveczky
in~\cite{fase10}. To support the definition of timed behaviors, this
approach provides a set of ``timed constructs,'' namely, timers and
clocks with different growth rates. These constructs are special Ecore
classes with a predefined Real-Time Maude semantics in MOMENT2. To
define timed behaviors, timer and/or clock objects must be added to
the models. In MOMENT2, adding timed behaviors can be done in a
\emph{non-intrusive} way (i.e., without modifying the original
metamodel) using MOMENT2's support for \emph{multi-model}
transformations~\cite{fase10}. 

The other  Eclipse-based framework for real-time systems with
Maude formal analysis support is the \emph{e-Motions} model
transformation
framework~\cite{Rivera-Duran-Vallecillo:2009,emotions}.  
One key difference between MOMENT2 and e-Motions is that the latter
puts even more emphasis on making models as high-level and intuitive as
possible. In contrast  to MOMENT2's textual
model transformation rules, e-Motions is designed to support the
definition of \emph{visual} DSMLs by specifying also a concrete syntax
for the elements in the metamodel, and by allowing 
the elements in the metamodel to map to graphical
objects. 

The approach to support timed behaviors in e-Motions is completely
different from the one in MOMENT2. Since the focus of e-Motions is on
providing a high-level intuitive formalism,
 e-Motions provides support for using timed
behaviors using different kinds of \emph{timed} model transformation
rules. 

MOMENT2 provides a built-in metamodel 
of basic timing constructs, thus allowing the possibility of associating
appropriate clocks and timers to the elements of user models, 
which are then manipulated by multimodel transformations.
To free the user from  the burden of dealing with  
timing constructs, e-Motions provides a high-level way of defining
timed behaviors by providing different kinds of \emph{timed model
transformation rules}. For example, model transformation rules can have a 
duration, which can be strict or given by a \emph{time interval}. 
Rules can be either \emph{atomic}, to represent actions that 
have no effect until it has been completed, or \emph{on-going}, to 
represent continuous actions. Rules can be declared periodic, 
eager or non-eager, etc. The different rules and their use is 
explained in more detail in Section~\ref{sec:eMotions}. 


Since e-Motions proposes such novel and powerful rules for defining
timed behaviors within a formal model transformation  framework, there
are two critical and closely related issues that must be addressed:
\begin{enumerate}
\item How can these new timed features be used to define ``real''
  timed systems and languages?
\item How convenient and intuitive is e-Motions in practice to
  define timed systems and languages in a visual style?
\end{enumerate}

Although e-Motions has already been used for developing several systems
(see, e.g., \cite{Rivera-Duran-Vallecillo:2009,Troya-Rivera-Vallecillo:2010}
and
\url{http://atenea.lcc.uma.es/E-motions}),
they were intended to illustrate 
 the e-Motions approach, and did not focus on timing aspects. 
This paper addresses the above issues by showing how e-Motions can be used to
visually specify and formally analyze systems in two very different
but prototypical classes of real-time systems: distributed 
network protocols, and scheduling
algorithms. 

Scheduling is a core issue in real-time  systems, and
plays a central role in modeling languages such as the AADL modeling
standard for avionics and automotive embedded systems used in
industry~\cite{aadl}. Obviously, if e-Motions should be able to handle
such embedded systems and AADL models, 
it must support the definition of scheduling in a simple and natural
way. 

Since e-Motions is a new framework with novel timed features we
want to illustrate, we focus on showing the entire specification of smaller
systems, instead of showing small isolated
fragments of large and complex systems. Therefore, in this paper, we
apply e-Motions to the following systems:
\begin{enumerate}
\item A simple protocol for measuring the \emph{round trip time}
  between  neighboring nodes in a network. Although this example is
  small,  it contains  many of the key features of
  distributed real-time protocols: local clocks and timers, nondeterministic
  message transmission times,\msgloss{ message loss,} periodicity of protocol
  runs, etc. An important  reason for selecting this example is also that it was
  used to illustrate timed model transformations in MOMENT2;
  therefore, 
  the reader can compare for herself the two very different styles of defining
  real-time systems in a formal model transformation framework. 
\item The well known \emph{earliest deadline first} scheduling
  algorithm (see,
  e.g.,~\cite{buttazzo}). 
\end{enumerate}

The paper is organized as follows: Section~\ref{sec:eMotions} briefly
recalls e-Motions; Section~\ref{sec:rtt} shows how the round trip
system has been specified and formally analyzed in e-Motions; and
Section~\ref{sec:edf} does the same for the scheduling
algorithm. Finally, Section~\ref{sec:concl} presents some concluding
remarks. 

\section{Modeling and Analyzing Time-Dependent Behavior with
  e-Motions} \label{sec:eMotions} 

e-Motions \cite{Rivera-Duran-Vallecillo:2009} is a DSML and a graphical tool
developed for Eclipse that extends in-place model transformation with a model
of time and mechanisms to state action properties, designed for the
specification of real-time DSMLs' behavior.

Once we have defined the structure of our language with a metamodel,
we are ready to define its dynamic behavior with e-Motions. The
e-Motions tool enables the use of a graphical concrete syntax
of the language in the specification of its behavior through the
definition of a graphical concrete syntax (\emph{gcs}) model (see \cite{rivera-thesis}). This
model is automatically generated from the definition of our DSML's
metamodel, and once it is created, we only have to assign a picture to
each metaclass of our metamodel.

One way of specifying the dynamic behavior of a DSML is by describing the
evolution of the modeled artifacts. In MDE, this can be
done using model transformations supporting in-place
updates~\cite{Czarnecki-Helsen:2003}. The behavior of the DSML is then specified
in terms of the permitted actions, which are in turn modeled by the model
transformation rules.
This approach provides a very intuitive way to specify
behavioral semantics, close to the language of the domain expert and at the right
level of abstraction~\cite{deLara:2006}.

The behavior of a DSML is then specified by a set of graphical in-place rules,
each of which
represents a possible \emph{action} of the system. These rules are of the form
$l:[\textrm{NAC}]^{*}\times\textrm{LHS}\rightarrow \textrm{RHS}$, where $l$ is
the rule's label (its name), and LHS (left-hand side), RHS (right-hand side)
and NAC (negative application conditions) are model patterns that represent
certain (sub-)states of the system. The LHS and NAC patterns express the
preconditions for the rule to be applied, whereas the RHS represents its
postcondition, i.e., the effect of the corresponding action. Thus, a rule can
be applied, i.e., triggered, if an occurrence (or match) of the LHS is found in
the model and none of its NAC patterns occurs. Generally, if several matches
are found, one of them is non-deterministically selected and applied, producing
a new model where the match is substituted by the appropriate instantiation of
its RHS pattern (the rule's \emph{realization}). The model transformation
proceeds by applying the rules in a non-deterministic order, until none is
applicable --- this behavior can be modified by some execution
control mechanism, e.g.,
strategies~\cite{maude-book,Rivera-Duran-Vallecillo:2009-simulation}. 


Contrary to standard in-place transformation approaches, the
e-Motions tool distinguishes two types of timed rules:
\emph{atomic} and \emph{ongoing} rules. One natural way to model
time-dependent behavior consists in decorating the
rules with their \emph{durations}, i.e., by assigning to each action
the time it takes. Atomic rules are defined in such a way. 
As normal in-place transformation rules, an atomic rule can be
\emph{triggered} whenever a match of its LHS, and none of 
its NAC patterns, is found in the model. Then, the action specified
by such rule is scheduled to be \emph{realized} 
between $t$ and $t'$ time units later, where $[t,t']$ represent the 
duration of the rules specified as an interval of time. At 
that time, the rule is applied by substituting 
the match by its RHS and performing the attribute
computations.
Atomic rules also admit a parameter that specifies the period of an
action. If a rule has this value set, the rule will be tried to be
triggered at the beginning of each period, and if it cannot be so
(i.e., if its precondition is not satisfied), it will not be enable
until the next period. 

Note that in e-Motions actions may have now a duration, and
therefore elements can be engaged in several actions at the same
time. 
The triggering of an atomic rule is only forbidden if another
occurrence of the same rule is already being executed \emph{with the same
participants}.\footnote{We call participants of a rule to those
elements that instantiate the rule's LHS pattern.}
The only condition for the final application of an atomic rule is that the 
elements involved are still there; otherwise the action will be \emph{aborted}. If
we want to make sure that something happens (or does not happen)
during the execution of an action, we can make use of \emph{action
  execution} elements to 
model the corresponding exceptional
behavior (see~\cite{Rivera-Duran-Vallecillo:2009} for further details). 

Ongoing rules allow the modeling of actions that are continuously progressing
and require to be continuously updated. Think for instance of an
action that models the behavior of local clocks, whose time increases
continuously 
with time.
These properties must be always updated, since other rules may depend
on their value. 
Ongoing rules are used to model actions that do not have \emph{a
priori} duration time --- they progress with time while the rule
preconditions (LHS and NACs) hold, or until the time limit of the
action (if set) is reached --- and are required to be continuously
updated --- their effects are always computed before the triggering of
any atomic rule.

Once we have defined the behavioral specifications of our DSML using
e-Motions, we can perform simulation, reachability analysis,
and LTL model
checking analysis of our DSML models. Currently, only simulation can be
performed directly in the e-Motions tool. In order to
perform reachability and model checking analysis we need to move to
the Maude system~\cite{maude-book} (which is also available for the Eclipse
platform).  In~\cite{emotions}, we show how
Maude can be  used to provide semantics to real-time DSMLs. 
The e-Motions tool automatically generates (by using model
transformations) the Maude specifications of the metamodel of
the DSML,
its behavior, and an initial  model. The result of the
transformation is a rewriting logic specification of the system, which
is executable and allows us to simulate and formally analyze it. 

In particular, the operator \verb+<<_;_>>+ is useful to define
analysis commands without having to know the Maude representation of an
e-Motions language. The term \verb+<<+ \emph{ocl-expr} \verb+;+
\emph{model} \verb+>>+ evaluates the OCL expression \emph{ocl-expr} in
the model \emph{model}. Therefore, to search for a model reachable
from an initial model \verb+myModel+ that satisfies the OCL expression
\verb+ocl-expr+, we can use the Maude search command 

\begin{lstlisting}[style=AMMA, language=Maude, numbers=none]
 search [1] init(myModel) =>* { MODEL:@Model } in time T:Time
   such that << ocl-expr ; MODEL:@Model >> .
\end{lstlisting}  

The generated models typically have an infinite number of reachable
states; to ensure that search commands terminate, one can either
restrict the search to a given depth (\verb+search [1,1000] ...+),
or, as proposed in \cite{rivera-thesis}, one can change  the
generated Maude specification by modifying the tick rule so that time
does not advance beyond a desired time bound.\footnote{This is slightly
  less convenient than in MOMENT2, where we can achieve the effect of
  \emph{time-bounded analysis} by just adding a new timer to the
  initial model.}

We can use the operator \verb+<<_;_>>+  to define atomic state
propositions for linear temporal logic (LTL) model checking
purposes. For example, to define a 
proposition \verb+p+ to hold in all models where the OCL expression
\verb+ocl-expr+ holds, we can define

\begin{lstlisting}[style=AMMA, language=Maude, numbers=none]
 op p : -> Prop [ctor] .
 ceq { MODEL:@Model } in time T:Time |= p = true if << ocl-expr ; MODEL:@Model >> .
\end{lstlisting}  

An LTL formula is then constructed from such atomic propositions and the
usual Boolean and LTL connectives, such as \verb+~+ (negation),
\verb+\/+ (disjunction), \verb+/\+ (conjunction), \verb+->+
(implication), \verb+[]+ (always), \verb+<>+ (eventually), \verb+U+
(until), and so on. One can then model check whether the LTL formula
\verb+formula+ holds in \verb+myModel+ by giving the Maude command

\begin{lstlisting}[style=AMMA, language=Maude, numbers=none]
 red modelCheck(init(myModel), formula) .
\end{lstlisting}  

\section{Specification and Analysis of a Round Trip Time Protocol}
\label{sec:rtt}

Estimating the \emph{round trip time} between two nodes in a network --- 
i.e., the time it takes for a message to travel from source to
destination, and back --- plays an important role in many large
communication protocols. In this
section we present a visual modeling language that can be used to
specify a simple round trip time estimation protocol. 
Section~\ref{sec:rtt-prot} gives a brief overview of the protocol,
Section~\ref{sec:rtt-spec} presents the modeling language specifying the
protocol, and Section~\ref{sec:rtt-anal} shows how the system can be
formally simulated and analyzed. 

\subsection{The Round Trip Time Protocol}
\label{sec:rtt-prot}

The round trip time protocol used in this paper is a fairly simple
protocol that aims to estimate the current round trip time between
neighboring nodes in a network as follows: Each node starts a round of
the protocol by sending a \emph{request} message to its (only)
neighbor\footnote{Note that the neighbor relation does not have to be
  symmetric.} with a time stamp that
records the local time at which the request was sent. When a node
receives a \emph{request} message, it immediately responds by sending
a \emph{response} message with the original time stamp back to the sender.
 When a node
receives the \emph{response} message containing its original time
stamp, it can easily  compute the round
trip time using its local clock. Since the network load may change,
and since messages may get lost, each node starts a new round of the
protocol \emph{every} 100 time units to get up-to-date round trip time
estimates. We assume that the message transmission time is between 5
and 20 time units, and that messages may be lost.

\subsection{The e-Motions Specification of the Protocol}
\label{sec:rtt-spec}

This section presents a DSML that defines the round trip time protocol
in e-Motions. 

The class diagram in Figure~\ref{fig:RTTPMM} shows the metamodel that
defines the abstract syntax of the DSML defining the protocol. A
\code{Node} object represents a node in the network, and is identified by
its \code{id} attribute. The \code{neighbor} attribute denotes the identifier of
its only neighbor\footnote{There are obviously many other ways to model
  such a system, including letting \textsf{neighbor} refer to a concrete
  object instead.} and  the \code{rtt} attribute denotes the desired current
round trip time estimate. The local time of a node is given by
the \code{time} attribute of the node's associated local clock. 
\figeps{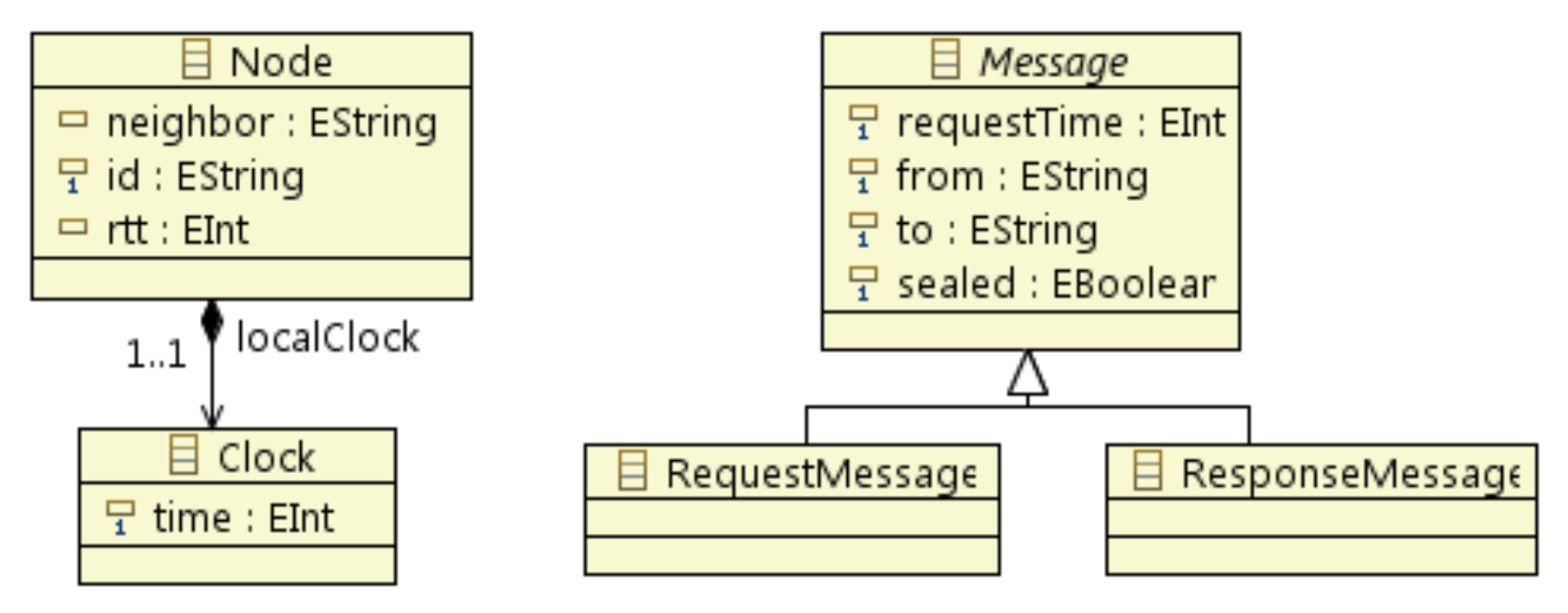}{48}{Abstract syntax of the round trip protocol language.}
We have two kinds of messages, \code{RequestMessage} and
\code{ResponseMessage}, 
that are subclasses of the generic \code{Message} class. A message contains
a \code{requestTime} attribute denoting the time stamp, and \code{to} and
\code{from} 
attributes  denoting, respectively, the receiver and sender
of the message. When a message is created, it is ``sealed,'' and the
message can only be read when the seal has been removed; this
models message delays as explained below.

The concrete syntax for the round trip time protocol language that
assigns a graphical object to each class in the abstract syntax is
given in Figure~\ref{fig:RTTPConcrete}.
\figeps{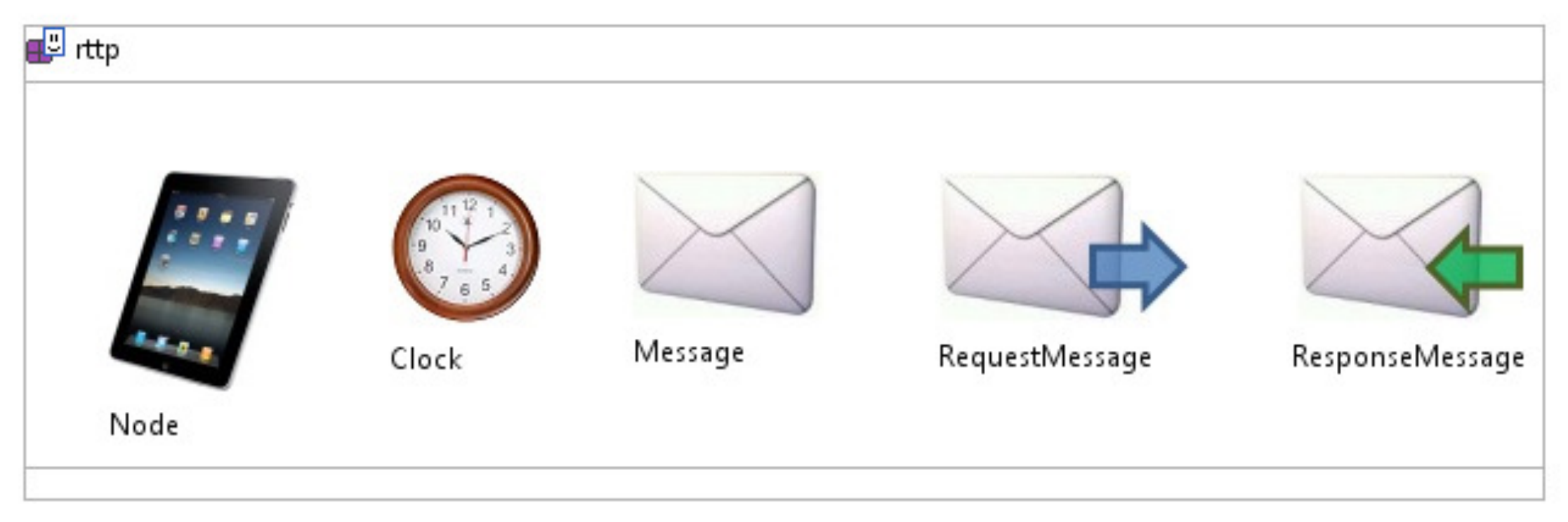}{48}{A concrete syntax for the round trip time
  protocol  language.}

The real-time behavior of the DSML can now specified by a set of
visual model transformation  rules. 

Figure~\ref{fig:RequestRule2} shows 
the \code{Request} rule, which starts a round of the
protocol for a node \code{n}. It models how the node \code{n} creates a
request message \code{m} to its neighbor. The message is initially
sealed, and includes the time of its creation (given by the node's
local clock). We  consider that the action is instantaneous, and
therefore the rule's  duration is set to the interval \code{[0,0]}.
\figeps{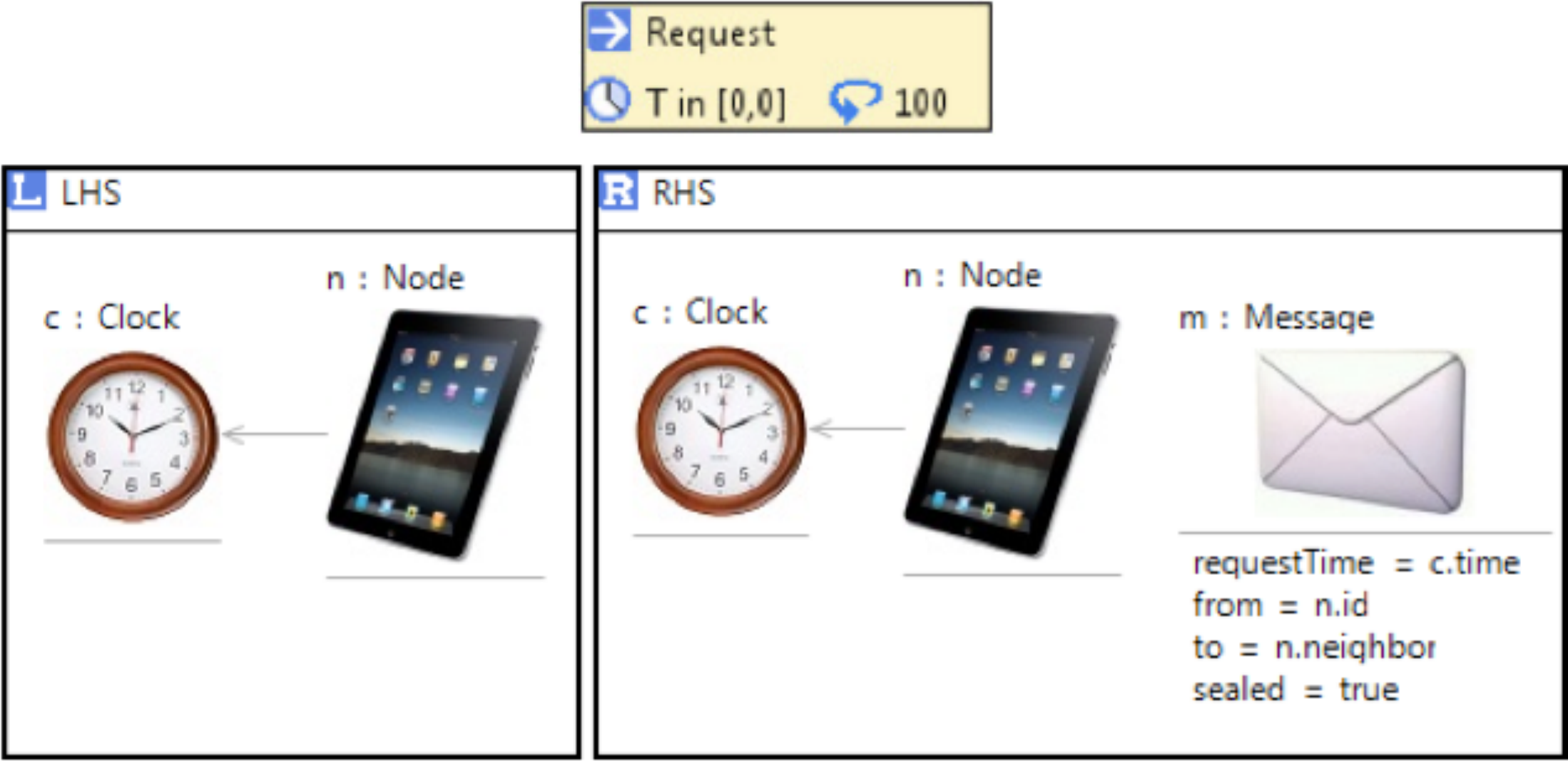}{50}{The \emph{Request} periodic rule.}
This is \emph{periodic} rule, with period 100 (notice the
loop icon in the header of the rule). 
Remember that, unless the rule is \emph{soft}, an execution of a rule is
scheduled as soon as possible for \emph{each} set of objects that
matches a rule's 
left-hand side. Therefore, the above rule will be applied every 100
time units \emph{for each
\code{Node}}.

Successful message transmission  is modeled by the \code{Transfer}
rule shown in Figure~\ref{fig:TransferRule}. 
\figeps{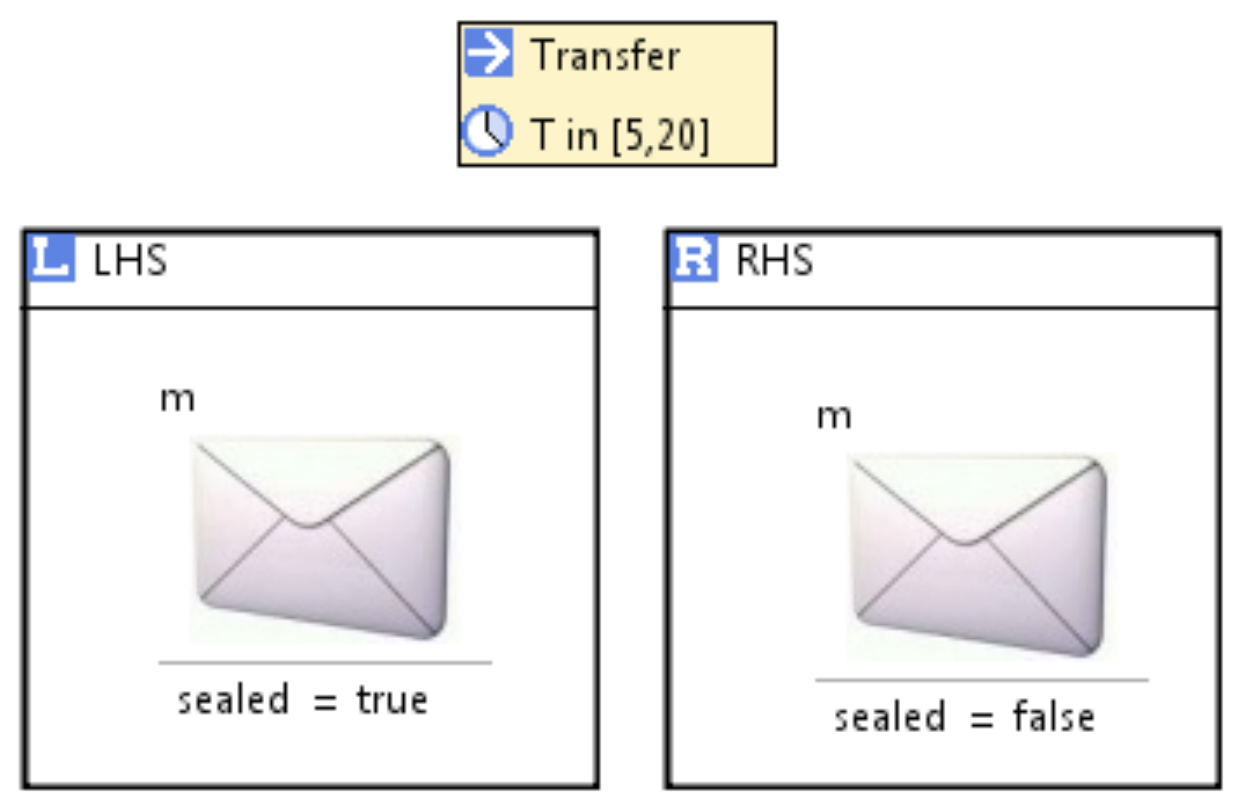}{52}{The \emph{Transfer} atomic rule.}
This action takes between five and twenty time units, and removes  the
seal of a message,  which indicates that the message  can be received. 
Any kind of message may get lost at any time, what is modeled by the \code{Lost} rule in Figure~\ref{fig:LostRule}.
\figeps{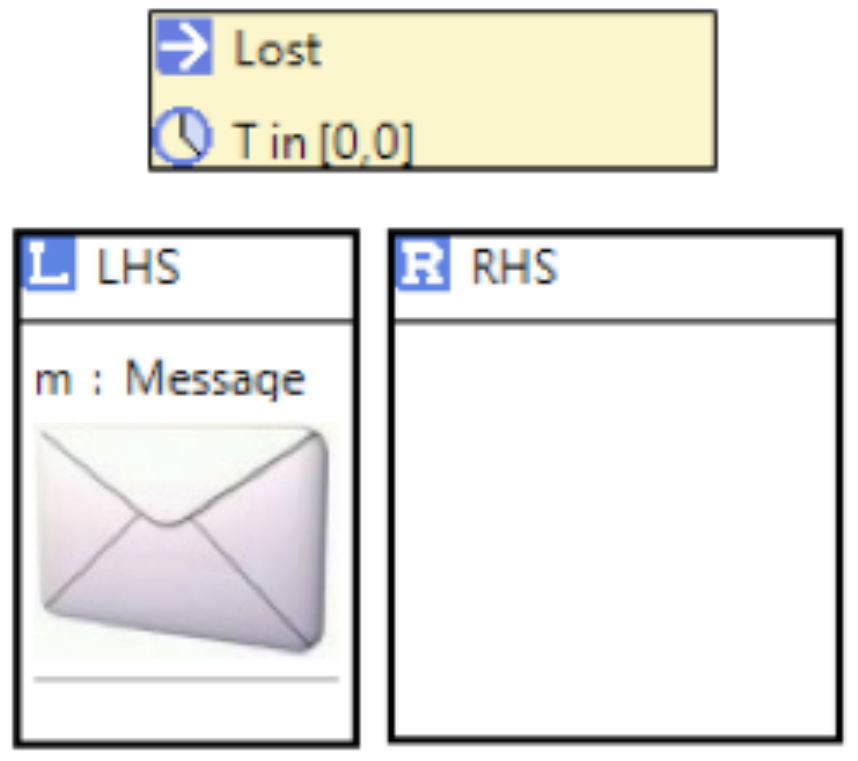}{40}{The \emph{Lost} atomic rule.}

Figure~\ref{fig:ResponseRule} shows the instantaneous atomic rule that
applies when the seal of a request message\footnote{Remember that
  messages with a (left to right) blue arrow are request messages, and messages
  with a (right to left) green arrow are response messages.} \code{m1} has been removed;
that is, the 
message is ready to be received. The intended recipient \code{n} then
consumes the message and sends a new (sealed) response message to the
sender of the request message. This new message  also includes the same
time
stamp that was received in the request message.
\figeps{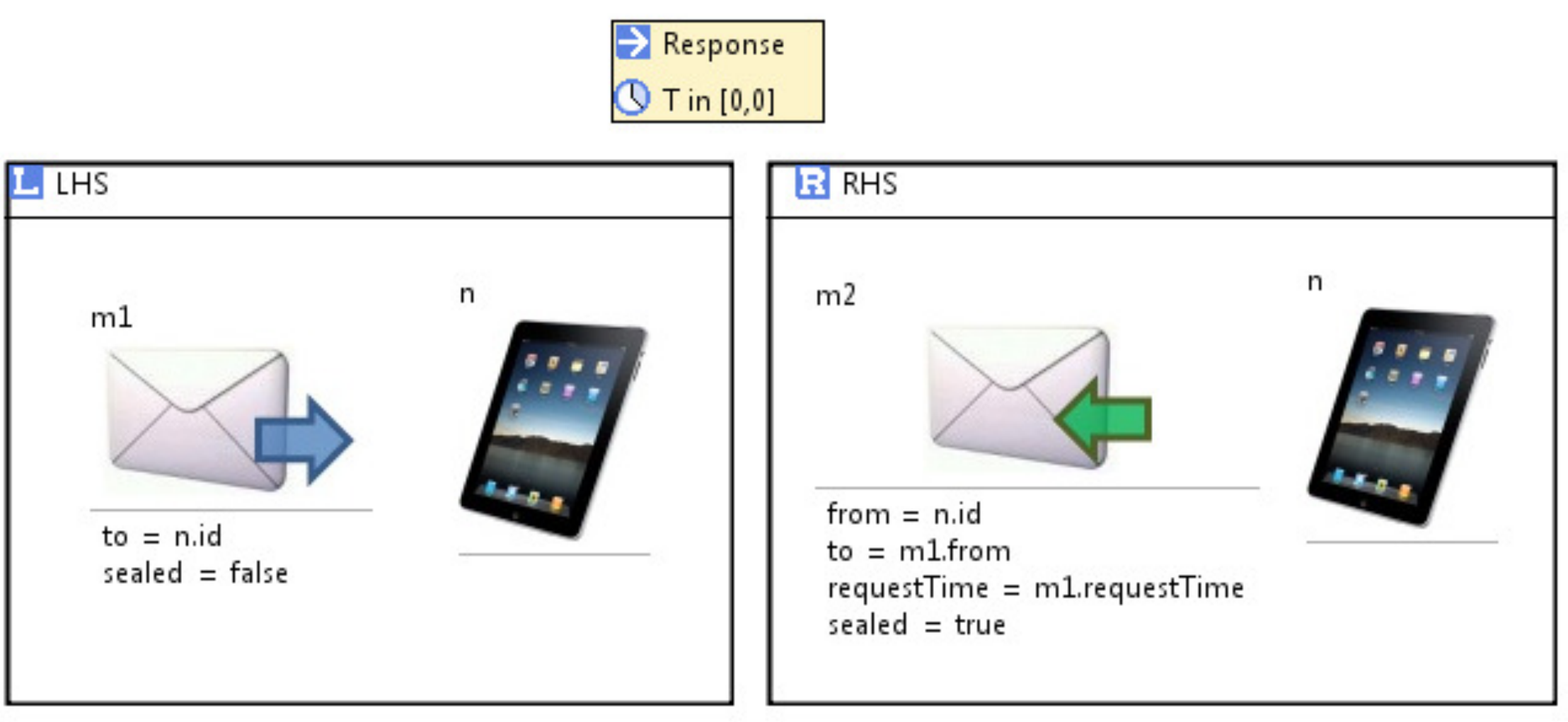}{70}{The \emph{Response} instantaneous rule.}
%

When  the response message is received, the instantaneous rule
\code{ComputeRtt} in Figure~\ref{fig:ComputeRttpRule} computes the receiver's 
\code{rtt} value by subtracting
the time stamp \code{m.requestTime} from the current value of its
local clock.
\figeps{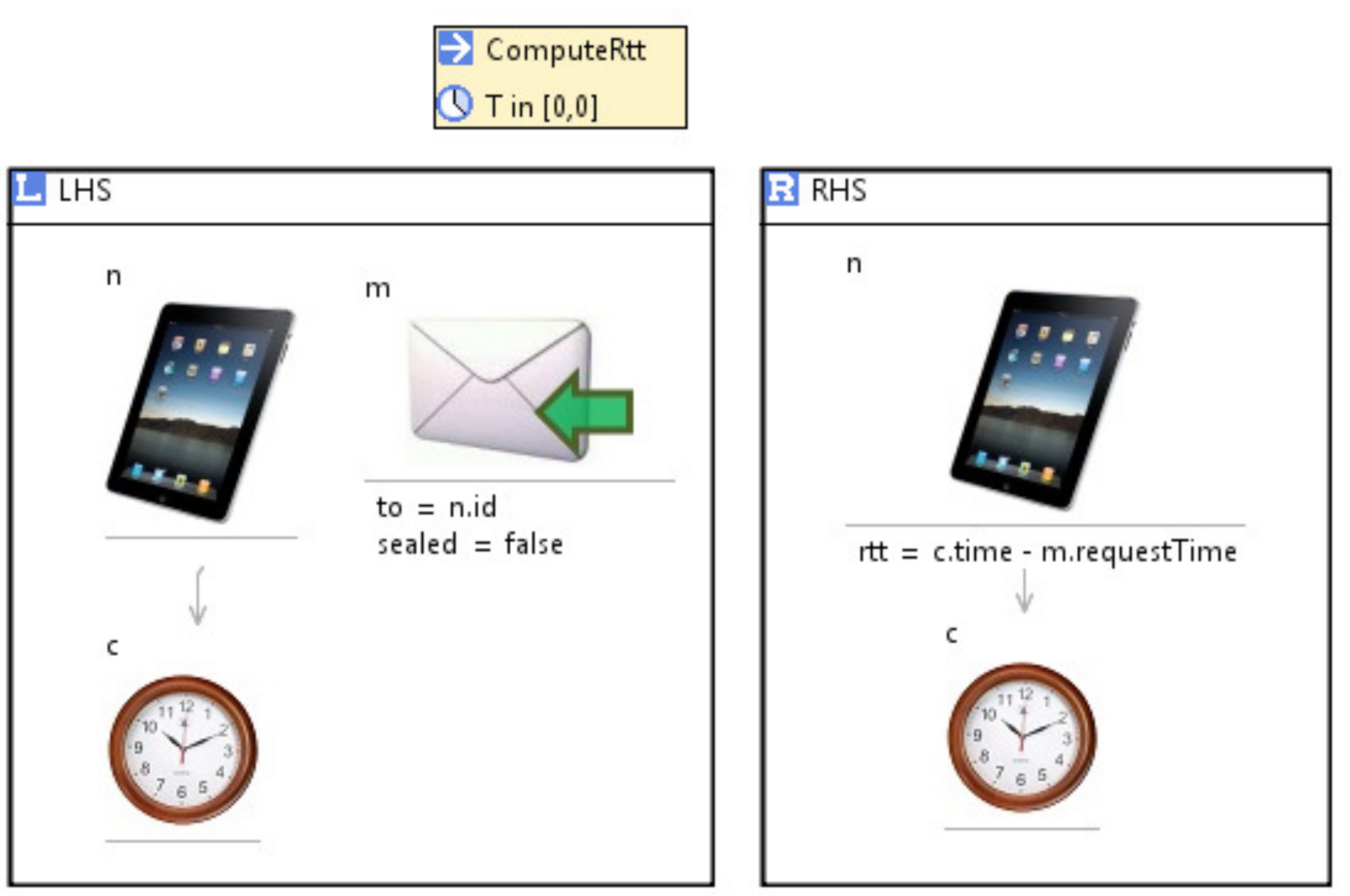}{75}{The \emph{ComputeRtt} instantaneous
  rule.}

Finally, the  \emph{ongoing} (notice the \emph{purple} arrow) rule \emph{LocalTimeElapse}, shown in
Figure~\ref{fig:LocalTimeElapseRule},  increases the \code{time}
value of each  clock in the system by  the elapsed time \code{T}. 
\figeps{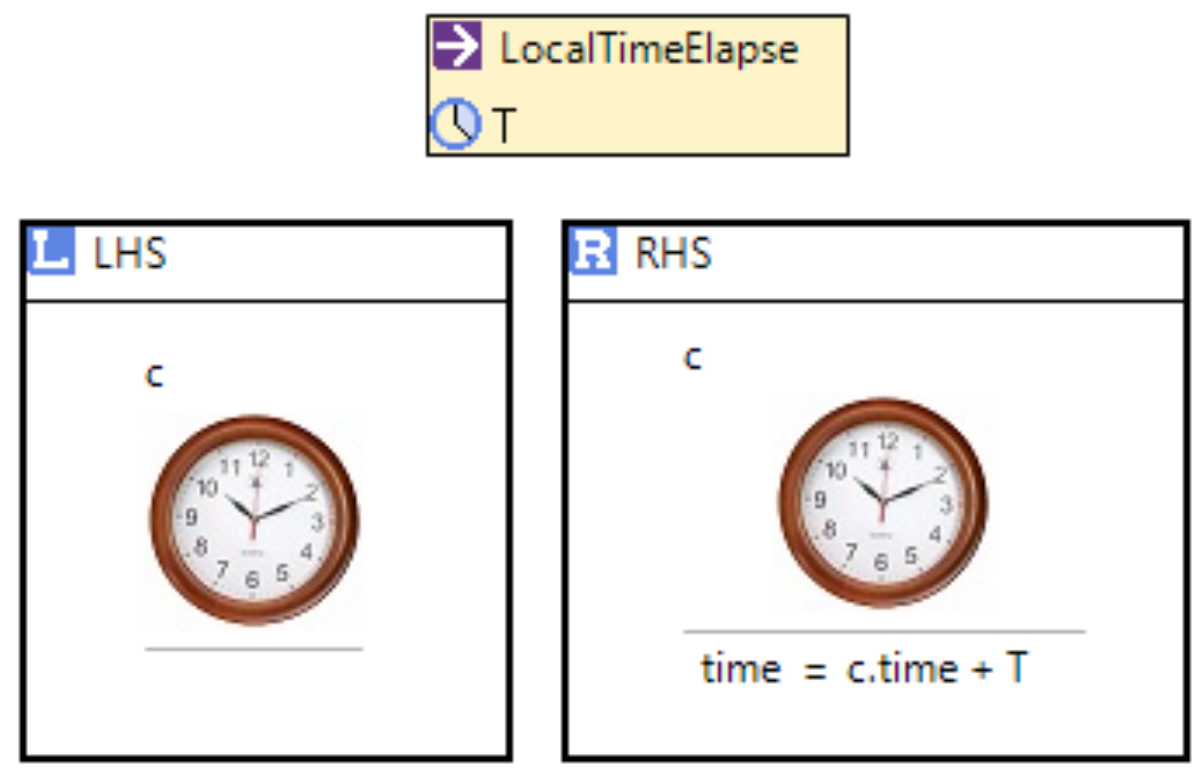}{56}{The \emph{LocalTimeElapse} ongoing rule.}

\subsection{Formal Analysis of the Protocol} \label{sec:rtt-anal}

Let the model shown in  Figure~\ref{fig:InitialModelRule}  be  the initial
model of the  system. This model is composed of two nodes and
their  local clocks.

\figeps{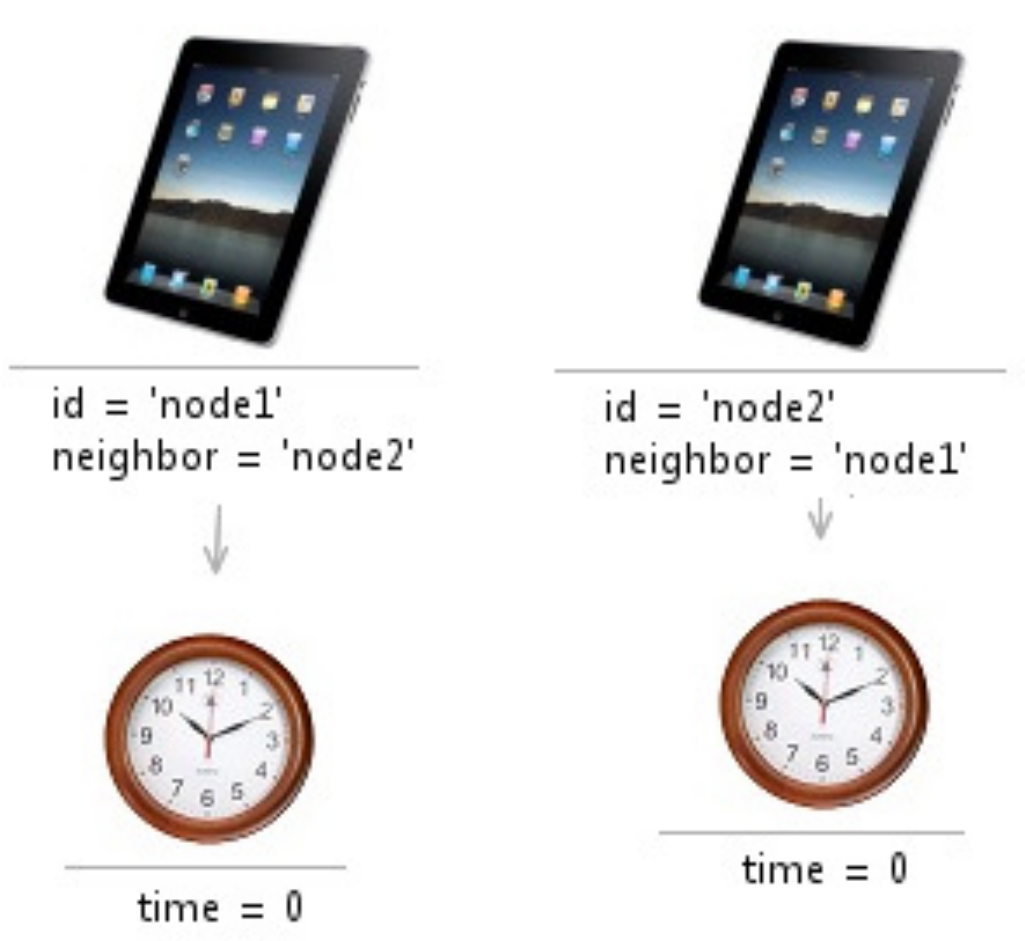}{60}{The \emph{rttpModel} initial model.}


Since a message transmission can take anywhere 
between five and twenty time units, we could expect that the
round trip times will range between ten and forty time units.
This property  can  easily be checked using the Maude's
\code{search} command. This command allows us to explore (following a 
breadth-first strategy) 
the reachable state space in different ways. Thus, we move to the
Maude environment, and  check if a model
with a node whose \code{rtt} attribute is set to a value lower than
ten time units or greater than forty time units can  be reached:

\begin{lstlisting}[style=AMMA, language=Maude, numbers=none]
 search [1] init(rttpModel) =>* { MODEL:@Model } in time T:Time
   such that << Node@rttp . allInstances 
                  -> exists ( n | (n . rtt@OCLSf > 40) or (n . rtt@OCLSf < 10)) ; 
                MODEL:@Model >> .

 No solution.
\end{lstlisting}  

With this command, we are looking for a model \code{MODEL} that
satisfies the condition expressed as an OCL expression in
the \code{such that} clause as explained in Section~\ref{sec:eMotions}. 
In this case, the search command does not find any solution, and
therefore we can conclude that starting from our initial model, the
round-trip time value of every node will always range between ten and
forty time units.  

We can also check whether the local clocks of the system are always
synchronized by searching for a state that does not satisfy it, i.e.,
by looking for a state on which two clocks have different time values: 
\begin{lstlisting}[style=AMMA, language=Maude, numbers=none]
 search [1] init(rttpModel) =>* { MODEL:@Model } in time T:Time
   such that << 'clock1 . time@OCLSf <> 'clock2 . time@OCLSf ; MODEL:@Model >> .

 No solution.
\end{lstlisting}

\section{Earliest Deadline First Scheduling} \label{sec:edf}

This section explains how we can visually specify and formally analyze
the well known \emph{earliest deadline first} (EDF) scheduling
algorithm in e-Motions. Even though model checking is not necessary to
decide schedulability of this simple protocol, for much more complex
scheduling algorithms where schedulability is very hard to analyze analytically, 
Real-Time  Maude model checking analyses can
indeed be successfully applied to  analyze
schedulability~\cite{fase06}. 

\subsection{The EDF Scheduling Algorithm}
 
In  EDF scheduling with periodic tasks (see,
e.g.,~\cite{buttazzo}), 
we are given one or more \emph{processors} and a set of tasks $\tau_1,
\ldots, \tau_n$, where each task $\tau_i$ is served by a constant
bandwidth \emph{server} $S_i$ with \emph{period} $p_i$ and
\emph{execution time}  $e_i$. The server $S_i$ executes the  instances
of $\tau_i$ in rounds of length $p_i$, and in each round it must have
access to a/the processor for time $e_i$ in total. 

Tasks can be \emph{preempted} (i.e., suspended in the middle of an
execution to allow the processor to execute tasks with higher
priority), and it is always the task(s) with the \emph{earliest
  deadline} that have the highest priority. That is, it is always the
server with the least amount of time remaining of its current round
that should be executing (unless it has ended its execution in the
round). 

Therefore, in EDF, the behavior of a server $S_i$ can be summarized as
follows: 
\begin{itemize}
\item At the beginning of each round: if a processor is idle, the
  server starts executing on the processor; if a processor is not
  idle, but the time until the end of the executing server $S_j$'s
  current  round is
  greater than the period $p_i$, then $S_i$ preempts (and suspends)
  $S_j$ and starts 
  executing; otherwise, the server $S_i$ starts waiting for an
  available processor.
\item  A waiting or suspended server may start executing when another
  server stops executing, and there are no other waiting/suspended
  servers with higher priority. 
\item When the server $S_i$ has executed for a total amount of time
  $e_i$ in its current round, it releases the processor, to the
  waiting/suspended server (if any) with the highest priority.
\item When a server's round  ends, it immediately starts a new
  round. 
\end{itemize}

\subsection{Specifying EDF in e-Motions}

The class diagram defining the abstract syntax of the DSML specifying
the EDF algorithm is given in Figure~\ref{fig:EDFMM}.
\figeps{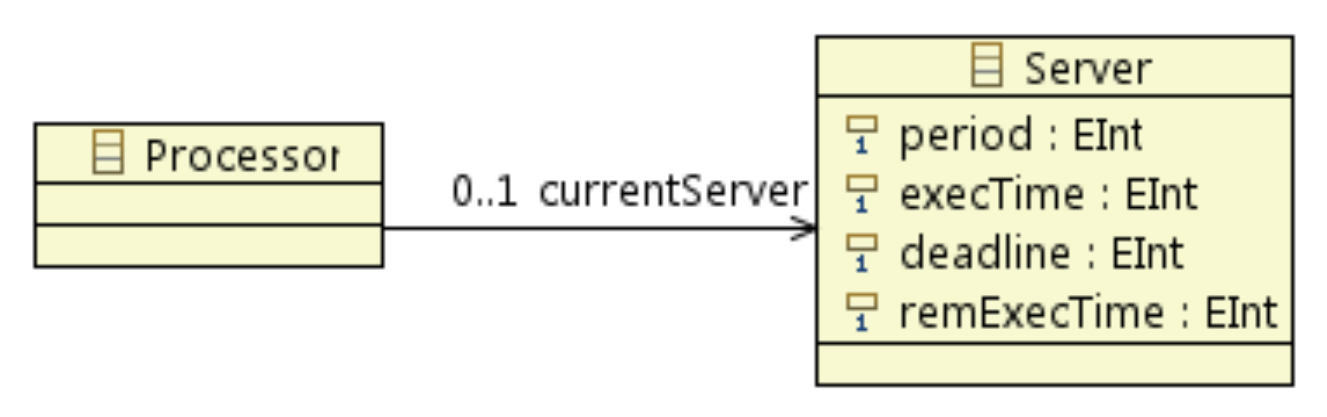}{30}{The EDF metamodel.}
In this metamodel, a system is composed of a set of servers and a set
of processors. A processor can
 execute at most a server at a time (see the \code{currentServer}
reference). Task servers  are identified by their period (attribute
\textsf{period}) and their execution time (attribute \code{execTime}).
The \code{deadline} and \code{remExecTime} attributes denote,
respectively,  the time until the end of the current round
and the remaining execution time in the current round. 
The visual concrete syntax we have defined for the EDF example is
shown in Figure~\ref{fig:EDFConcrete}.
\figeps{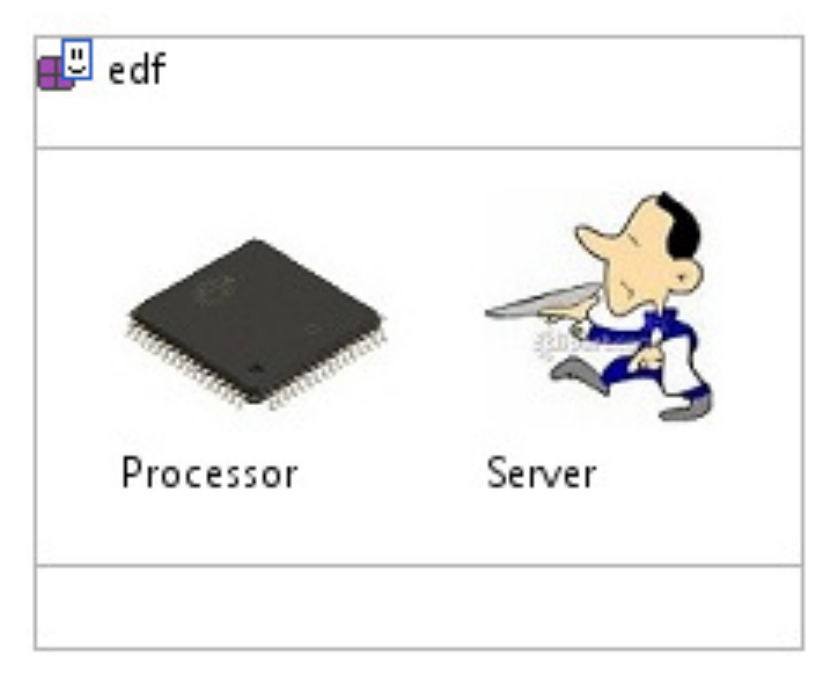}{51}{A concrete syntax for the EDF language.}

The  model transformation rules in Figures~\ref{fig:NewPeriodRule2}-\ref{fig:DecreaseDeadlineRule} define the dynamic behavior
of EDF. 

When the current round is over (i.e, the time remaining
until the \code{deadline} is 0), the server \code{s} starts a new
round by first setting its \code{deadline} attribute to the length of
its period, and by setting  the value of its \code{remExecTime}
attribute, that denotes the
remaining execution time of the new round,  to whatever what left of
the execution time in the previous round plus the execution time
needed in this round.\footnote{If the system is schedulable, then
  \textsf{s.remExecTime} is always 0 at the end of each period.}
This is modeled by the instantaneous atomic model
transformation rule \emph{NewPeriod}  in Figure~\ref{fig:NewPeriodRule2}.
\figeps{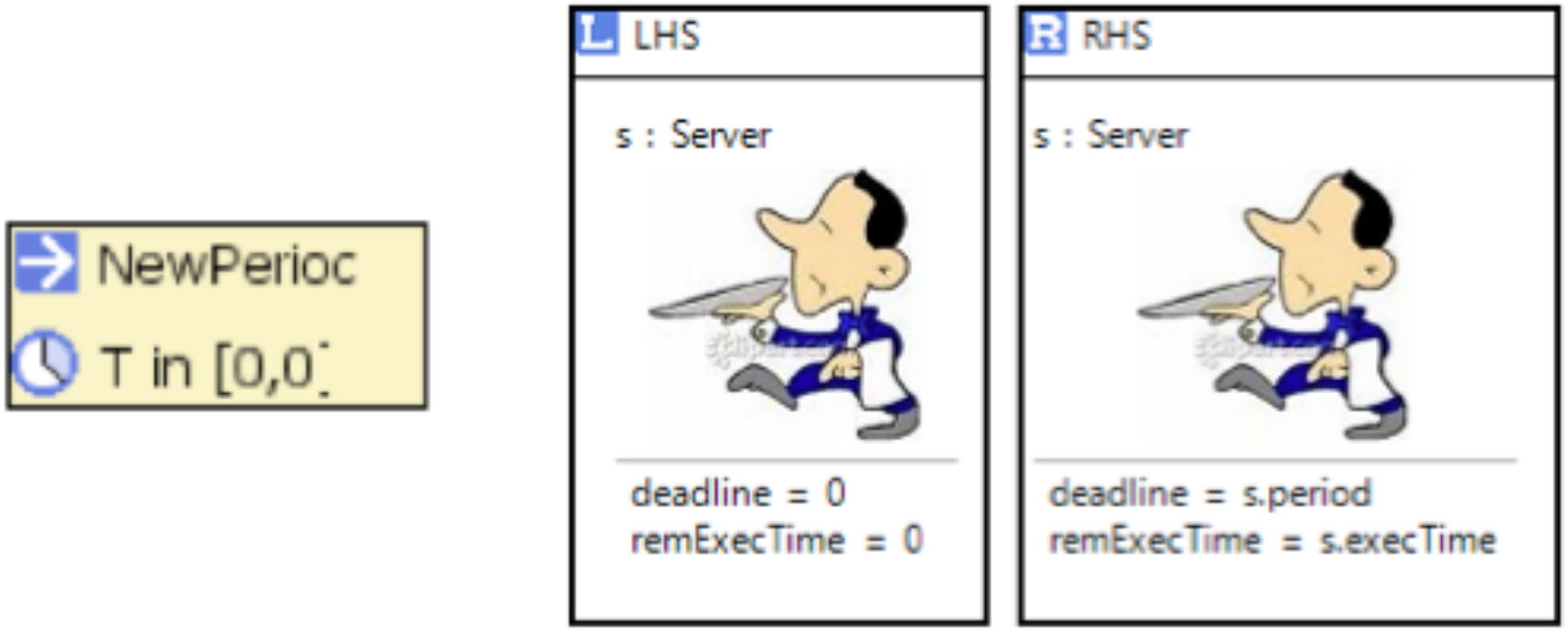}{40}{The \emph{NewPeriod} instantaneous rule.}
As already mentioned, unless an atomic rule is declared to be
\emph{soft}, it is applied as soon as it becomes
enabled. 

In the  instantaneous atomic rule \emph{ServerSelection},
shown in Figure~\ref{fig:ServerSelectionRule},  the
server \code{s} gets to execute \emph{if} there is no other server
with remaining execution in the period with earlier deadline, \emph{and} the
currently executing server (if any) has a strictly later deadline than
\code{s}. 
\figeps{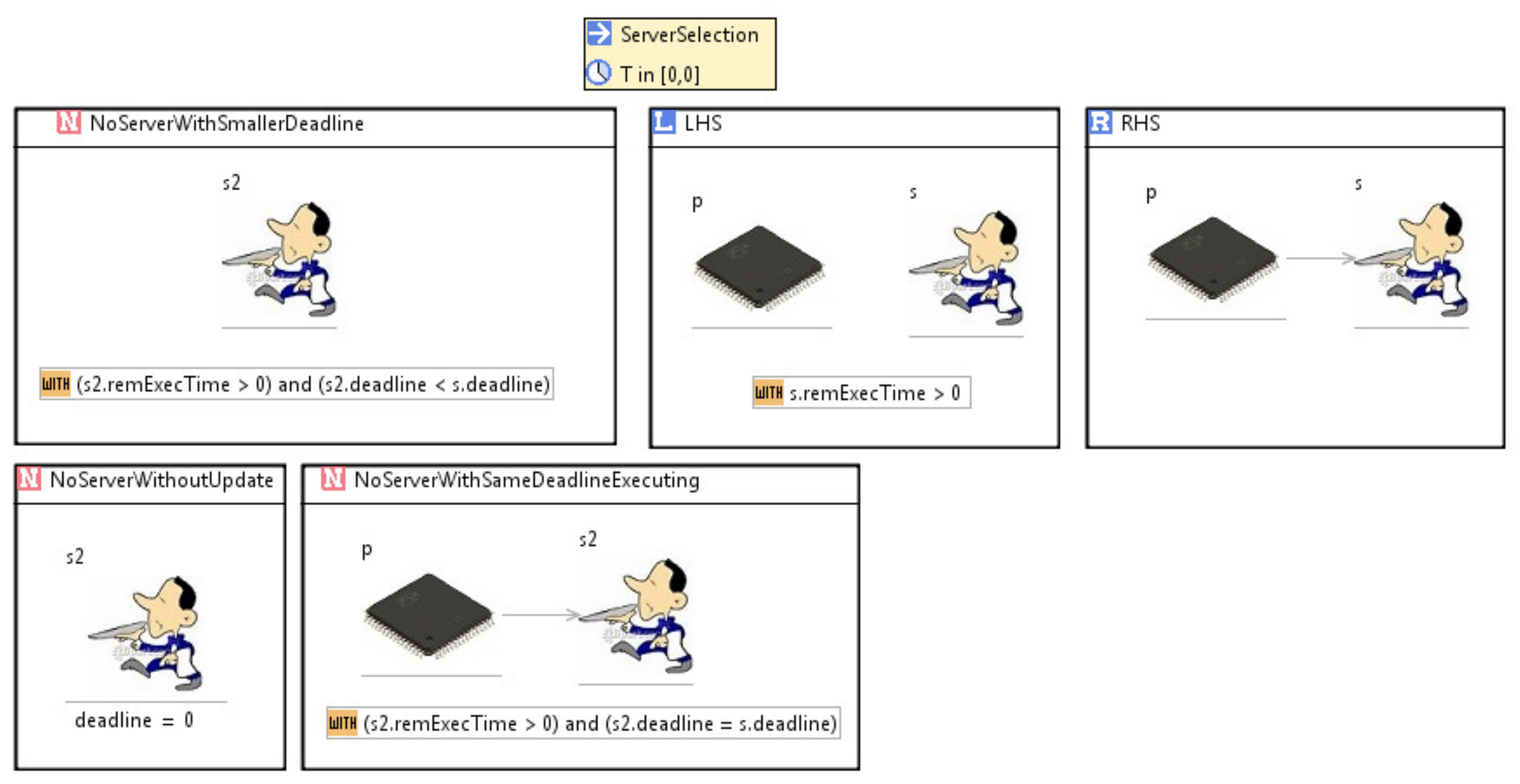}{80}{The \emph{ServerSelection}
  instantaneous rule.} 

The rule \emph{ServerSelection} also  applies
when a server \code{s2} finishes its execution in a round, if
there is a waiting/suspended server. The case when  a server \code{s}
finishes its execution (i.e.,
\code{s.remExecTime} is 0), but there is no server waiting, is modeled
by the  instantaneous atomic rule \emph{ReleaseProcessor} in
Figure~\ref{fig:ReleaseProcessorRule2},  where
\code{s} just releases the server at the end of its current execution
if no server is waiting. 
\figeps{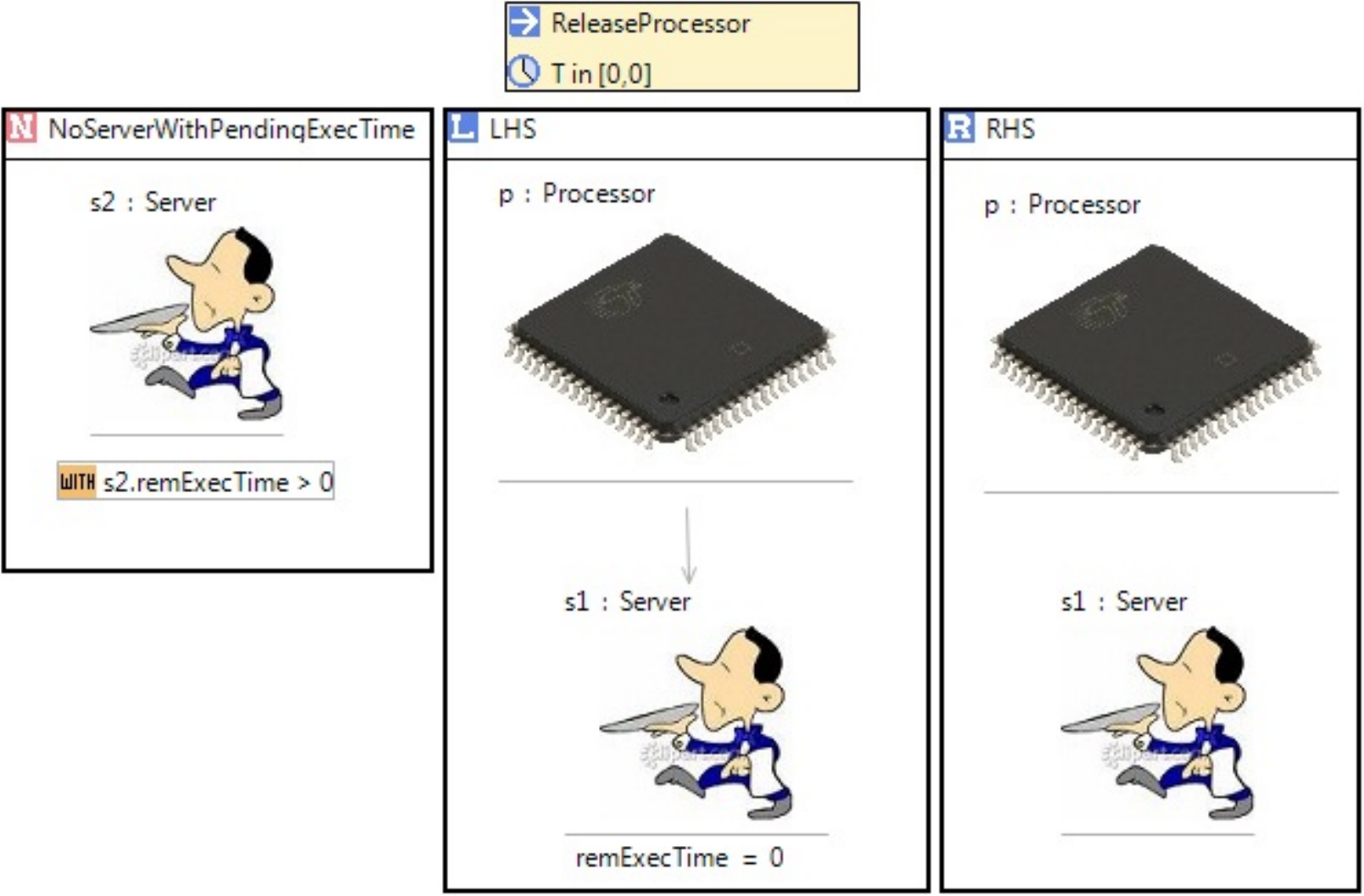}{70}{The \emph{ReleaseProcessor}
  instantaneous rule.}

Finally, we have two \emph{ongoing} rules: Rule \emph{Execution} in
Figure~\ref{fig:ExecutionRule} decreases the remaining deadline of each
server according to the elapsed time \code{T}, until the node reaches
the end of its round, and the rule \emph{DecreaseDeadline} in Figure~\ref{fig:DecreaseDeadlineRule} decreases
the remaining execution time of each \emph{executing} server according to
the elapsed time.

\figeps{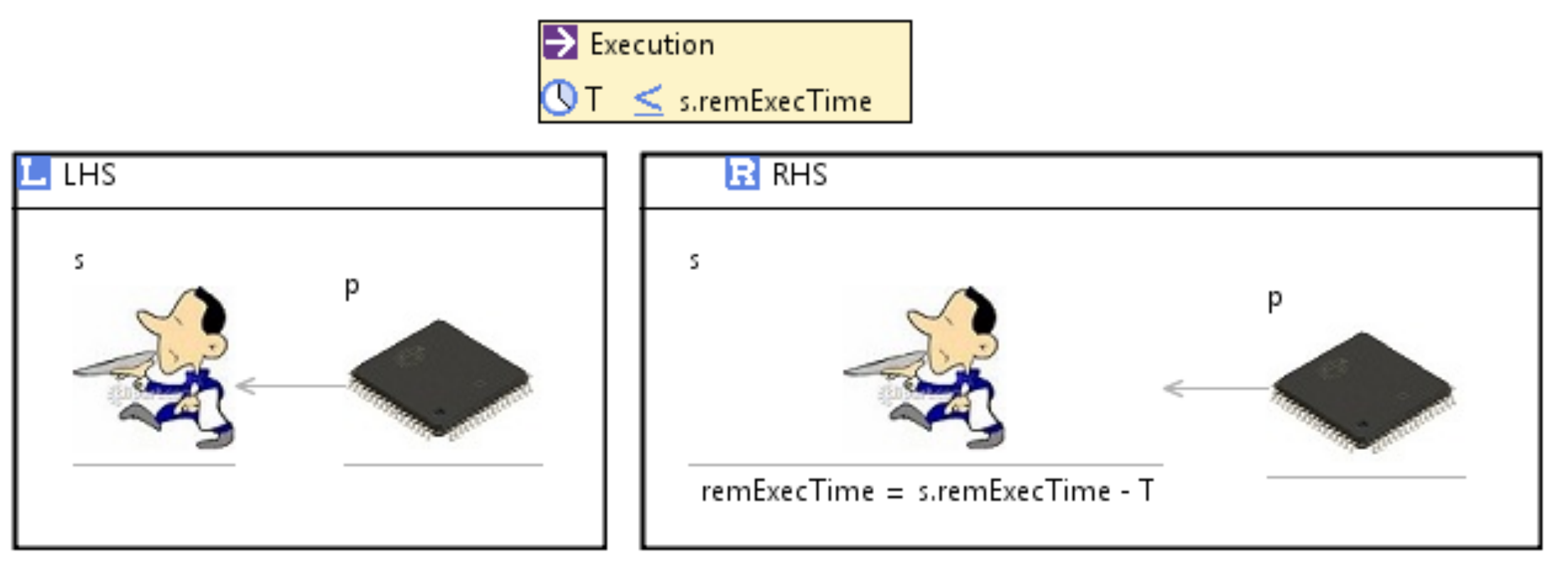}{56}{The \emph{Execution} ongoing rule.}

\figeps{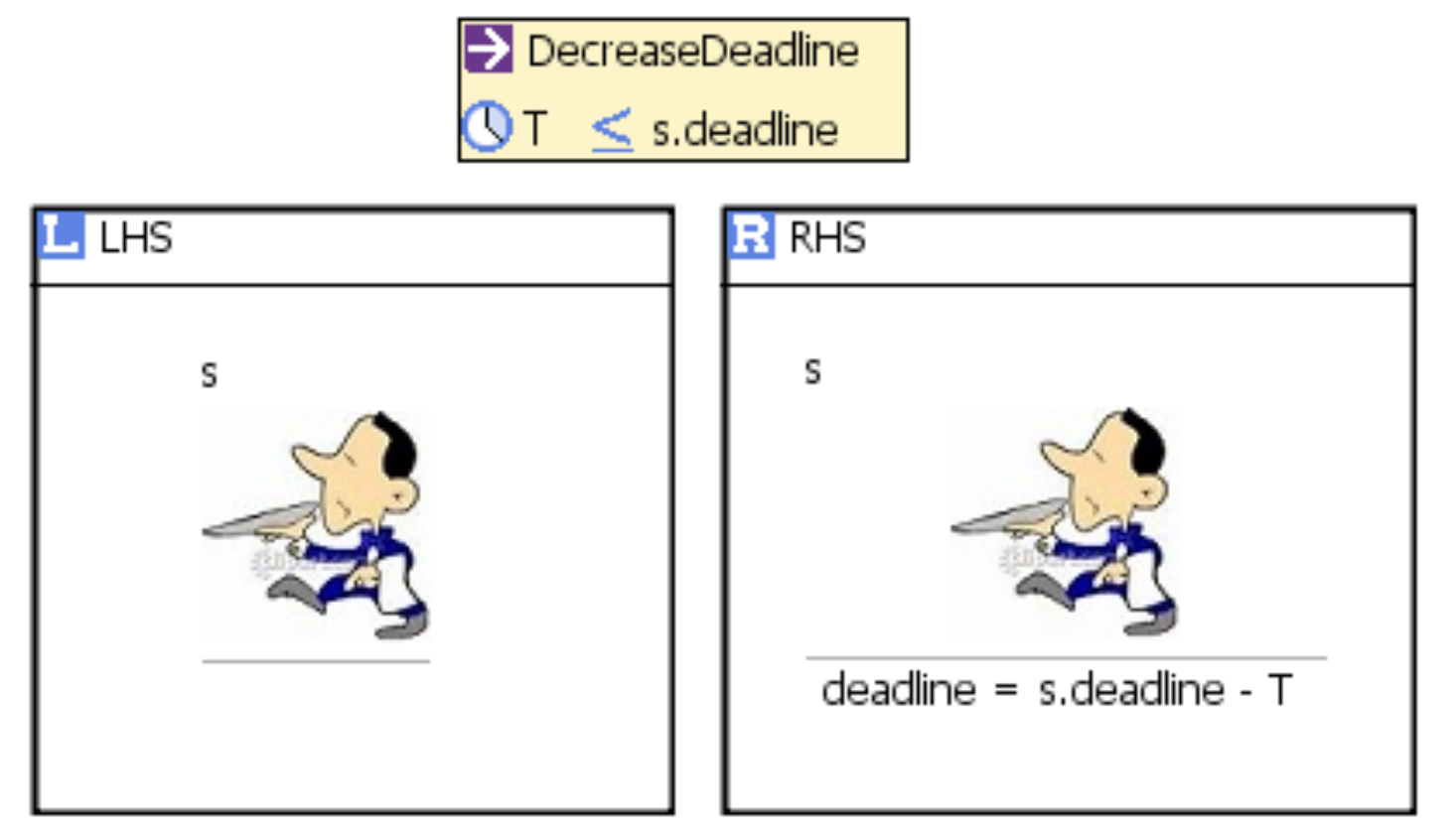}{56}{The \emph{DecreaseDeadline} ongoing rule.}

\subsection{Analyzing the EDF Algorithm}
\label{sec:anal-edf}

We can use the generated Maude specification of the EDF language to
analyze whether a given set of servers is schedulable; that is, whether
all servers can execute for the allocated execution time
\code{execTime} in 
each round. 

\figeps{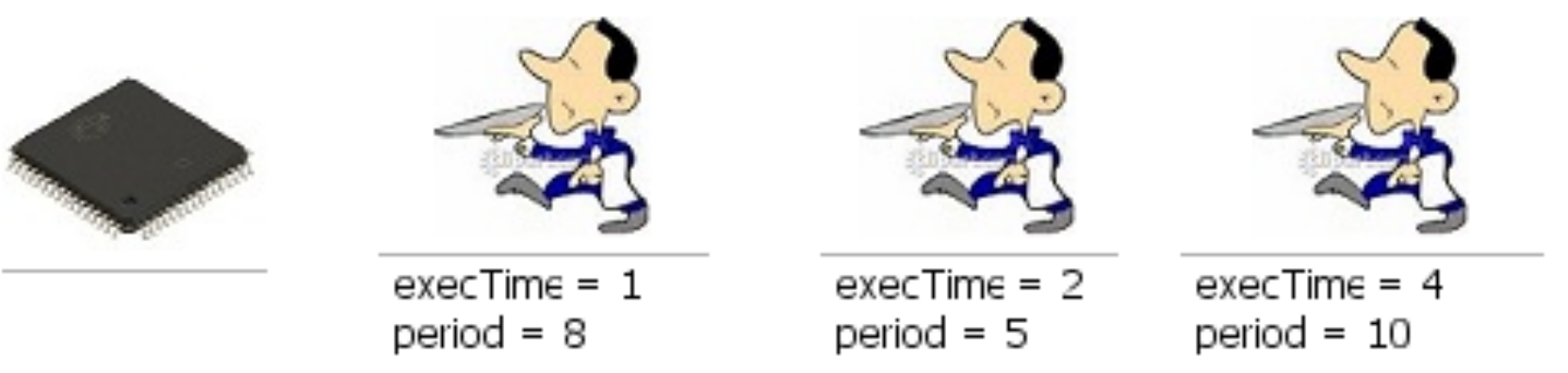}{25}{The \emph{edfModel} initial model.}
Figure~\ref{fig:InitialModelEDF} shows an initial model
\verb+edfModel+ with three 
servers and one processors. To analyze whether this server set  is 
schedulable, we  search for a reachable model which can \emph{not}
execute for the desired amount of time in some round; that is, a model
in which the  remaining execution time is
greater than the deadline for some server \verb+s+:

\begin{lstlisting}[style=AMMA, language=Maude, numbers=none]
 search [1] init(edfModel) =>* { MODEL:@Model } in time T:Time
  such that << Server@edf . allInstances 
              -> exists ( s | s . deadline@OCLSf < s . remExecTime@OCLSf ) ; MODEL:@Model >> .	
\end{lstlisting}

\section{Concluding Remarks}\label{sec:concl}

We have shown how two fairly small, but ``non-artificial,'' real-time
systems in very different domains can be specified in the visual model
transformation framework e-Motions. We have also shown how
the Maude tool can be used  to formally analyze the 
automatically synthesized Maude models. We leave it up to the reader
to evaluate the convenience of using e-Motions  for the
specification and analysis of real-time systems, and also encourage
her to compare this work with the specification and analysis of the
same 
round trip example in the timed model transformation framework
MOMENT2~\cite{fase10}. 

Our personal impression is that it  seems quite convenient to model
systems in this way; however,  the analysis support could be better
integrated within e-Motions, despite the very significant advantage of
having the \verb+<<_;_>>+ operator, which allows us to easily define
any reachability and LTL model checking query without knowing anything
about  the internal Maude representation  of our system.

\bibliographystyle{eptcs} 
\bibliography{bibl}

\end{document}